**Simulation-based Estimation of Mean and Standard Deviation for Meta-analysis via Approximate Bayesian Computation (ABC)**

Deukwoo Kwon[1*]
*Corresponding author
Email: DKwon@med.miami.edu

Isildinha M. Reis[1,2]
Email: ireis@med.miami.edu

[1]Sylvester Comprehensive Cancer Center, University of Miami, Miami, FL 33136
[2]Department of Public Health Sciences, University of Miami, Miami, FL 33136

**Abstract**

**Background:** When conducting meta-analysis of a continuous outcome, estimated means and standard deviations from the selected studies are required in order to obtain an overall estimate of the mean effect and its confidence interval. If these quantities are not directly reported in the publications, they need to be estimated from other reported summary statistics, such as median, minimum, maximum, quartiles.

**Methods:** We propose a simulation-based estimation approach using Approximate Bayesian Computation (ABC) technique for estimating mean and variance based on various sets of summary statistics found in published studies. We conduct simulation study to compare the proposed ABC method with existing methods by Hozo et al. (2005), Bland (2015), and Wan et al. (2014).

**Results:** In estimation of standard deviation, our ABC method performs best in skewed or heavy-tailed distributions. The average relative error (ARE) approaches zero as sample size increases. In normal distribution, our ABC performs well. However, Wan et al. method is best since it is based on normal distribution assumption. When distribution is skewed or heavy-tailed, ARE of Wan et al. moves away from zero even as sample size increases. In estimation of mean, our ABC method is best since AREs converge to zero.

**Conclusion:** ABC is a flexible method for estimating study-specific mean and standard deviation for meta-analysis especially with underlying skewed or heavy-tailed distributions. The ABC method can be applied using other reported summary statistics such as posterior mean and 95% credible interval when Bayesian analysis is employed.

**Background**

In medical research, it is common to conduct systematic review and meta-analysis to provide an overall estimate of a clinical treatment outcome from a set of individual studies. When the outcome is continuous, in order to conduct meta-analysis we need estimated means and the corresponding standard deviation (or equivalently, variances) from the selected studies. However, not all studies report these quantities directly. Instead, studies may report mean and confidence interval, p-value, median, minimum and maximum, range or interquartile range (IQR). As another example, when Bayesian methods are employed in data analysis, posterior means and 95% credible intervals are usually reported.

If mean and standard deviation are not directly reported in the publication, these need to be estimated from the other reported summary statistics. Wiebe et al. [1] describe several methods, including algebraic and approximate algebraic recalculations, to obtain standard deviation estimate from confidence levels, t-test or F-test statistics, and p-values. Based on descriptive statistics (such as median, minimum and maximum, range or IQR), the *ad-hoc* approach is a study-level imputation. For instance, the sample median is often used as the estimate of the sample mean assuming symmetric distribution, and the sample standard deviation is commonly estimated by either $\frac{range}{4}$ or $\frac{IQR}{1.35}$.

Hozo et al. [2] proposed a simple alternative method for estimating the sample mean and the sample standard deviation from the median, minimum, maximum, and the size of the sample. Another alternative method was proposed by Bland [3] estimating these quantities based on minimum, first quartile, median, third quartile, maximum, and sample size. The methods by Hozo et al. [2] and by Bland [3] make no assumption on the distribution of the underlying data. Recently, Wan et al. [4] proposed a method that improved estimation of the sample standard deviation based on median, minimum, maximum, and the size of the sample. Wan et al. [4] also provided a method for estimating standard deviation based on median, quartiles, and the size of the sample. Their method is based on ordered statistics and assumes normal distribution of outcome measure.

In this paper, we propose an Approximate Bayesian Computation (ABC) approach for estimating mean and standard deviation which produced more precise estimates of true parameters as sample size increases and it accommodates various distributions.

In Section 'Methods' we summarize the methods by Hozo et al. [2], Bland [3] and Wan et al. [4] and describe our proposed ABC method. In 'Results', we describe and report the finding of the simulation studies comparing the performance of these methods. We use the statistical software R in performing all statistical programming related to the implement of the various methods, analysis and simulations.

## Methods

We denote the sample descriptive statistics as follows: minimum ($x_{min}$), first quartile ($x_{Q1}$), median ($x_{med}$), third quartile ($x_{Q3}$), maximum ($x_{max}$), and the sample size (n). Let also consider the following three scenarios of available summary statistics. The first scenario (S1) assumes available only minimum, median, maximum and sample size (S1 = {$x_{min}$, $x_{med}$, $x_{max}$, n}). The second scenario (S2) assumes additionally having estimates of the first and third quartiles (S2 = {$x_{min}$, $x_{Q1}$, $x_{med}$, $x_{Q3}$, $x_{max}$, n}). The third scenario (S3) assumes having median, first quartile, third quartile, and sample size (S3 = {$x_{Q1}$, $x_{med}$, $x_{Q3}$, n}).

*Method of Hozo et al.*

Hozo et al. [2] proposed the following formulas for estimating mean and variance under S1= { $x_{min}$, $x_{med}$, $x_{max}$, n}:

$$\bar{x} \approx \begin{cases} \frac{x_{min}+2x_{med}+x_{max}}{4} & n \leq 25 \\ x_{med} & n > 25 \end{cases} \quad (1)$$

and

$$S^2 \approx \begin{cases} \frac{1}{12}\left(\left(\frac{x_{min}+2x_{med}+x_{max}}{4}\right)^2 + (x_{max} - x_{min})\right), & n \leq 15 \\ \left(\frac{x_{max}-x_{min}}{4}\right)^2 & 15 < n \leq 70 \\ \left(\frac{x_{max}-x_{min}}{6}\right)^2 & n > 70 \end{cases} \quad (2)$$

Hozo et al. approach implies different formulas for estimating mean and variance depending on sample size n. When sample size is between 26 and 70, Hozo et al.'s formulas in equations (1) and (2) are exactly the same as mean and variance formulas by the *ad-hoc* approach.

*Method of Bland*

Bland [3] extended the method of Hozo et al. by adding first quartile ($x_{Q1}$) and third quartile ($x_{Q3}$) to S1. That is, Bland's method provides formulas to estimate mean and variance under S2 = {$x_{min}$, $x_{Q1}$, $x_{med}$, $x_{Q3}$, $x_{max}$, n}. While Hozo et al. used the sample size to decide the formula to be employed in estimating mean and variance, the method by Bland incorporates the sample size in the proposed formulas:

$$\bar{x} = \frac{(n+3)x_{min} + 2(n-1)(x_{Q1} + x_{med} + x_{Q3}) + (n+3)x_{max}}{8n}$$

$$= \frac{(x_{min} + 2x_{Q1} + 2x_{med} + 2x_{Q3} + x_{max})}{8} + \frac{3(x_{min} + x_{max}) - 2(x_{Q1} + x_{med} + x_{Q3})}{8n}$$

$$\approx \frac{(x_{min}+2x_{Q1}+2x_{med}+2x_{Q3}+x_{max})}{8} \qquad (3)$$

and

$$S^2 = \frac{(n+3)(x_{min}^2 + 2x_{Q1}^2 + 2x_{med}^2 + 2x_{Q3}^2 + x_{max}^2) + 8(x_{min}^2 + x_{max}^2)}{16n}$$

$$+ \frac{(n-5)\left(x_{Q1}(x_{min} + x_{med}) + x_{Q3}(x_{med} + x_{max})\right)}{n}$$

$$\approx \frac{(x_{min}^2+2x_{Q1}^2+2x_{med}^2+2x_{Q3}^2+x_{max}^2)}{16} + \frac{x_{Q1}(x_{min}+x_{med})+x_{Q3}(x_{med}+x_{max})}{8} - \bar{x}^2. \qquad (4)$$

Note that in equation (4), the third term is the squared value of mean estimate using equation (3). As pointed by Wan et al., the 2$^{nd}$ term including sample size (n) can be dropped when sample size is large. Thus, after dropping the second term in (3), the estimator in (4) is independent of the sample size (n). Wan et al. proposed alternative estimators under S2, as described in next section 2. 3.

### *Method of Wan et al.*

Wan et al. [4] main focus was on improvement of standard deviation estimation. They proposed estimation formulas for the mean and standard deviation under the three scenarios, S1, S2, and S3, of available summary statistics.

For estimation of mean, Wan et al. proposed in S1 the same formula (1) by Hozo et al.[2], and in S2 the same formula (3) by Bland [3]. In S3 = { $x_{Q1}$, $x_{med}$, $x_{Q3}$, n}, they proposed the following new estimation formula for mean:

$$\bar{x} \approx \frac{(x_{Q1}+x_{med}+x_{Q3})}{3}. \qquad (5)$$

For estimation of standard deviation, Wan et al. proposed the following formulas:

$$S \approx \frac{(x_{max}-x_{min})}{2\Phi^{-1}\left(\frac{n-0.375}{n+0.25}\right)} \quad \text{in S1}, \qquad (6)$$

$$S \approx \frac{1}{2}\left(\frac{(x_{max}-x_{min})}{2\Phi^{-1}\left(\frac{n-0.375}{n+0.25}\right)}\right) + \frac{1}{2}\left(\frac{(x_{Q3}-x_{Q1})}{2\Phi^{-1}\left(\frac{0.75n-0.125}{n+0.25}\right)}\right) \quad \text{in S2,} \tag{7}$$

$$S \approx \frac{(x_{Q3}-x_{Q1})}{2\Phi^{-1}\left(\frac{0.75n-0.125}{n+0.25}\right)} \quad \text{in S3.} \tag{8}$$

Note that the standard deviation estimator in S2, equation (7), is simply the weighted average of those in S1 and S3, per equations (6) and (8), respectively. Wan et al. estimator of standard deviation is based on normality assumption and uses approximation of expected values of the order statistics.

*Simulation-based method via Approximate Bayesian Computation (ABC)*

We propose a simulation-based method using Approximate Bayesian Computation (ABC) technique to estimate sample mean and standard deviation.

Bayesian inference needs likelihood function as well as priors for parameters in the model. Given a likelihood function, f(θ|D), where θ denotes parameter of interest and D denotes observed data, and prior distribution, p(θ), on the parameter space, Θ, our statistical inference is based on posterior distribution of θ, p(θ|D)∝f(θ|D)p(θ). In some situations, likelihood function is analytically or computationally intractable. In meta-analysis, we combine selected studies with respect to a certain clinical outcome. However, the datasets of these studies are usually not accessible. Although we can construct likelihood function based on the probability model, we cannot evaluate likelihood function due to unavailability of all data points. Using the Approximate Bayesian Computation (ABC) approach, the likelihood can be replaced by a comparison of summary statistics from the observed data and those from simulated data using a distance measure. The ABC methodology was introduced by Tavaré et al. [5] in population genetics using a simple rejection algorithm in order to avoid the computation of the likelihood function via a simulation from a specific distribution. Marin et al. [6] provided extensive review of ABC methods.

Table 1 describes how to use ABC method for estimation of mean and standard deviation using summary statistics. The first step is to generate a set of candidate values for parameters, $\theta^*$, from a specific prior distribution, $p(\theta)$. The second step is to generate pseudo data, $D^*$, from the likelihood function $f(\theta^*)$. The third step is to decide whether $\theta^*$ is accepted or not. This decision depends on the distance between summary statistics of the observed data, $S(D)$, and those of simulated data, $S(D^*)$, i.e., $\rho(S(D),S(D^*))$, where $\rho(\bullet,\bullet)$ denotes a distance measure such as Euclidean distance used in our application of ABC. If $\rho(S(D),S(D^*))$ is small than a fixed tolerance value $\varepsilon$ (i.e., $\rho(S(D),S(D^*))<\varepsilon$) $\theta^*$ is accepted, otherwise it is rejected. Steps 1-3 are repeated a large number of times (e.g., N=20,000) in order to obtain multiple sets of $\theta^*$ for the inference. The fundamental idea of ABC is that a good approximation of the posterior distribution can be obtained using summary statistics and a small tolerance value, $\varepsilon$. Instead of setting small value for $\varepsilon$, we can also use the acceptance percentage. For example, with acceptance percentage of 0.1% and N=20,000, we select $\theta^*$ corresponding to the top 0.1% with smallest distance.

Thus, in order to apply ABC algorithm to estimate mean and standard deviation using reported descriptive statistics, the first step is to choose a distribution to be used for generating data. Given a set of descriptive statistics and the nature of outcome variable, an educated decision about the distribution can be made. For example, if clinical outcome is some score of health-related quality of life (e.g. The Expanded Prostate Cancer Index Composite (EPIC) score ranging from 0 to 100), then such a variable is bounded and in this case we can use beta distribution. For unbounded variable we can choose either normal distribution or log-normal distribution. When variable is change between two measurements, normal distribution is a good choice. When variable is either percentage or strictly positive, then log-normal, exponential, or Weibull are good choices. Prior distribution is determined by corresponding distribution in the previous step. If normal or log-normal distribution is chosen we need to specify two parameters, $\mu$ and $\sigma$. For Weibull distribution, shape and scale parameters are needed. For beta distribution, two shape parameters are needed. Usual choice of the prior is uniform distribution with relative wide range. When a chosen distribution belong to location-scale family, we can use an educated guess for location parameter, $\mu$. Instead of uniform distribution with huge range, we can use given descriptive statistics such as minimum ($x_{min}$) and first quartile ($x_{Q1}$) (maximum ($x_{max}$) and third quartile ($x_{Q3}$)) for lower bound (upper bound) of uniform distribution. Other prior distributions for shape and scale parameters are uniform between zero and some large number.

The estimates of mean and standard deviation by ABC are obtained based on accepted parameter values. For instance, when we consider normal distribution, average of accepted values for $\mu$ is the estimated mean; likewise, the average of accepted values for $\sigma$ is the estimated standard deviation.

For other distributions, estimates of mean and standard deviation can be obtained from plug-in method or simulation. Both approaches give comparable estimates. The plug-in method consists of replacing means of accepted parameter values into the corresponding formulas for mean and standard deviation. For example, in beta distribution, mean is $\alpha/(\alpha+\beta)$ and variance is $\alpha\beta/[(\alpha+\beta)^2(\alpha+\beta+1)]$. We obtain estimates of mean and standard deviation by replacing in these formulas $\alpha$ and $\beta$ with mean of accepted values for these parameters.

The simulation approach consists of obtaining mean and standard deviation from simulated samples using each set of accepted parameter values. For example, in beta distribution, given a set of accepted values of α and β, we generate pseudo data of same sample size and calculate mean and standard deviation in pseudo data. We repeat this procedure for all sets of accepted parameter values. The simulation estimates of mean and standard deviation are the average of means and average of standard deviations, respectively.

Although we can begin with an educated guess for determining distribution for ABC, we often face multiple distributions for our choice. For instance, outcome is related to distribution with positive support, there are several distributions to be considered, such as, log-normal, Weibull, or exponential. In this situation we rely on model selection (i.e. distribution selection in our context) while we apply ABC method. Bayesian model selection is usually based on either Bayes factor or marginal posterior probability of model. Let $M_1$ and $M_2$ be two models according to two different distributions (e.g., normal and beta distributions). Bayes factor is define as

$$B_{12} = \frac{P(M_1|D)/P(M_2|D)}{P(M_1)/P(M_2)}, \tag{9}$$

where $P(M_i)$ is the prior and $P(M_i|D)$ is the marginal posterior distribution of model $M_i$, i=1,2, and D denotes data. When we assume that $P(M_1)=P(M_2)=0.5$ then Bayes factor is a ratio of two marginal posterior distributions of model, $P(M_1|D)/P(M_2|D)$. In the ABC approach, data is not available we replace summary statistics, S, for D. The Bayes factor and marginal posterior probability of model can be approximated by acceptance frequency for each model (i.e., distribution) in the ABC. It can be extended when we consider more than two distributions to compare. An example of distribution selection is given in 'Discussion'.

# Results

*Designs of simulation studies*

In order to facilitate comparison between our ABC method and existing methods, the parameters of our simulation studies were set to be similar to that by Hozo et al. and Wan et al. for the three different scenarios of available descriptive statistics.

Under S1, we compare ABC to Hozo et al. and Wan et al. Under S2, we compare ABC, Bland and Wan et al. methods. And under S3, we compare ABC and Wan et al. methods. In addition, we examine the effect of skewness in estimation performance using log-normal and beta distributions.

Under S1, we use the same five distributions which both Hozo et al. and Wan et al. simulated: normal distribution with mean 50 and standard deviation 17, N(50,17); log-normal distribution with location parameter=4 and scale parameter=0.3, LN(4,0.3); Weibull distribution with shape parameter=2 and scale parameter=35, Weibull(2,35), beta distribution, Beta(9,4); and exponential distribution with mean=10, Exp(10).

Under S2, we use log-normal distribution with same location parameter value of 5 and three different scale parameter values (0.25, 0.5, and 1) in order to evaluate effect of skewness. We also use three beta distributions, Beta(5,2), Beta(1,3), and Beta(0.5,0.5), to examine effect of skewness and bimodality in estimation for bounded data distribution.

Under S3, we use the four distributions in S1 to investigate further the effect the choice of descriptive statistics for the standard deviation estimation.

In each scenario we consider 10 sample sizes (n= 10, 40, 80, 100, 150, 200, 300, 400, 500, 600). We obtain a sample of n observations from a particular distribution, and compute the true sample mean and standard deviation. Using the different methods (Hozo et al. Bland, Wan et al. and ABC) we obtain the various estimates of mean and standard deviation from the corresponding descriptive statistics. The relative errors (REs) are calculated as follows:

$$RE\ of\ mean = \frac{(estimated\ \bar{x} -\ true\ \bar{x})}{true\ \bar{x}}, \qquad (9)$$

and

$$RE\ of\ standard\ deviation = \frac{(estimated\ S -\ true\ S)}{true\ S}. \qquad (10)$$

For each sample size n, we repeat this procedure 200 times to obtain average relative errors (AREs).

In the simulations, we set acceptance percentage 0.1% and 20,000 total number of iterations for ABC method. Hence, we obtain 20 accepted parameter values for a specific distribution. Prior setting for each distribution in the ABC model for the simulation is described in Table 2.

### *Results of simulation studies*

In the simulation studies we compare estimation performance of the various methods in terms of average relative error (ARE) for estimating mean and standard deviation. In the next three subsections we present comparison of methods for standard deviation estimation. In last subsection, we present comparison among methods for mean estimation.

### *Comparison of Hozo et al., Wan et al., and ABC in S1 for standard deviation estimation*

In Figure 1 we show AREs in estimating standard deviation for the 3 methods as function of sample size under simulated data from the selected five distributions. The corresponding densities are displayed in figures 1A (normal, log-normal, and Weibull), 1E (beta) and 1G (exponential). Under normal distribution (Figure 1B) in S1 (that is, when $x_{min}$, $x_{med}$, $x_{max}$, n are available), while Hozo et al. method (solid square linked with dotted line) shows large average relative errors for sample size less than 300, Wan et al. method (solid diamond linked with dashed line) shows quite good performance over all sample sizes. The ABC method (solid circle linked with solid line) shows decreasing error as sample size increases, with AREs close to that for Wan et al. method for n ≥ 80.

Under log-normal distribution (Figure 1C), Hozo et al. method shows better performance between sample size of 200 and 400. Wan et al. method still shows good performance, though there is a tendency of AREs moving away from zero as sample size increases. ABC method has slightly worst performance than Wan et al. method when sample size is less than 300, and it is the best when sample size is greater than 300, and it is the worst for small sample size around n=10.

For Weibull data (Figure 1D), the ABC method is the best, showing very small AREs close to zero over all sample sizes. Wan et al. method clearly shows that ARE moves away from zero as sample size increases.

For data from beta or exponential distributions (Figure 1F and 1H), ABC the method performed best, showing AREs approaching zero as sample size increases. Wan et al. method shows opposite tendency of increasing ARE as sample size increases.

*Comparison of Bland, Wan et al., and ABC in S2 for standard deviation estimation*

In this simulation we compare estimation of standard deviation under these methods in S2 (that is, $x_{min}$, $x_{Q1}$, $x_{med}$, $x_{Q3}$, $x_{max}$, n are available) and examine the effect of violation of normality using log-normal distribution. We consider three log-normal distributions with same location parameter value 5 but three different scale parameters (Figure 2A). For LN(5,0.25),

Wan et al. and ABC methods have similar small average relative error. Bland's method shows largely underestimate in small sample size, and average relative error keep increasing as sample size increase. Note that AREs become overestimated when sample size is over 200. As data are simulated from more skew to the right distributions (Figures 2C and 2D), we see large estimation errors in Bland and Wan et al. methods. Wan et al. method shows increasing ARE as sample size increases, while Bland understate in small sample size n and overestimate in large n. The AREs of ABC method are large with small sample size when skewness increases; however, ARE of ABC method becomes smaller and approaches zero as sample size increases.

We also examine performance of these methods when data are simulated from three beta distributions. (Figure 3) In this simulation study, we investigate the effect of bimodality as well as skewness for bounded data. For all methods underestimation is depicted, with ABC performing best for n>40. Under skewed distribution (Figures 3B and 3C) Bland and ABC methods show the same pattern, however ABC shows much better performance since ARE approaches zero with increasing sample size. When underlying distribution is bimodal (Figure 3D), all three methods show large underestimation, although ABC continues performing best for n> 40 showing smaller AREs.

*Comparison of Wan et al. and ABC in estimating standard deviation under S1, S2, and S3*

Here we simulate data in S1, S2 and S3 and under four distributions: log-normal, beta, exponential, and Weibull distributions. In Figure 4, crossed symbols denote S1, open symbols S2, and solid symbols S3. Circle and diamond denotes ABC method and Wan et al., respectively. Under the several distributions, AREs for ABC method converge toward zero as sample size increases for the 3 scenarios, while Wan et al. fail to show this pattern.

*Comparison of methods for mean estimation*

We compare AREs for mean estimation between Wan et al. and ABC methods. Note that mean formula is the same between Wan et al.[4] and Hozo et al.[2] under S1, and between Wan et al. and Bland[3] under S2. Figure 5 indicates that our ABC method is superior in estimating mean when sample size is greater than 40 for all scenarios. Under log-normal in S1 the pattern of ARE of mean estimate for ABC in S1 is similar to that of standard deviation estimate for ABC (see Figure 1C). However, as sample size increases ARE approaches to zero.

## Discussion

The main factor that has a huge influence in the performance of the three methods is the assumed parametric distribution; especially when the samples are drawn from a skewed heavy-tailed distribution. Since inputs for the estimation of standard deviation in S1 are minimum value ($x_{min}$), median ($x_{med}$), and maximum value ($x_{max}$), the two extreme values vary a lot from data to data. The bad performance of ABC method under normal, log-normal and exponential distributions with small sample size can be explained by erratic behavior of two extreme values as an input. However, as sample size increases, ARE of ABC method becomes small and ABC is better than the other methods. Wan et al. method is based on normal distribution assumption. Thus, it performs well under the normal distribution or any distribution close to symmetric shape (e.g., beta(4,4) is symmetric at 0.5). When underlying distribution is skewed or heavy-tailed, although Wan et al. method incorporates sample size into the estimation formulas, ARE keep deviating from zero as sample size increases.

As we mentioned earlier, in order to perform ABC we need to choose an underlying distribution model, which it can be based on an educated guess. However the choice can be the distribution with the highest marginal posterior model probability among several candidate distributions. We performed a small simulation to see how this approach is reliable for selecting appropriate distribution for ABC. We generate samples of size 400 from beta(9,4). We compute marginal posterior probabilities of model for beta, $P(M_1|D)$, and for normal, $P(M_2|D)$. Note that $P(M_2|D)=1- P(M_1|D)$, when only two distributions are considered. We repeat 200 times to get how many times beta distribution is chosen, as well as to get the estimates of average of marginal posterior model probabilities. The beta distribution was chosen 157 times among 200 repeats (78.5%), average of $P(M_1|D)$ was 0.63 and average $P(M_2|D)$ was 0.37. The AREs of estimated standard deviation using beta and normal distributions were -0.0216 and 0.0415, respectively. The ARE of estimated mean using beta distribution was 0.00068 and it is quite smaller than that of normal distribution (0.0118). These results indicate that the distribution section procedure works well.

In our simulation for ABC method, we set acceptance percentage of 0.01% and N= 20,000 iterations. Computation using these values takes less than a minute. In real application we suggest to use N=50,000 or more iterations to get enough number of accepted parameter values for estimating mean and standard deviation.

In this paper we implement the ABC method using simple rejection algorithm. Other algorithms available include Markov chain Monte Carlo (ABC-MCMC; Marjoram et al.) and sequential Monte Carlo (ABC-SMC; Toni et al.). In future research, we plan to explore these methods for improving estimation of mean and standard deviation.

## Conclusion

We propose a more flexible approach to estimate mean and standard deviation for meta-analysis when only descriptive statistics are available. Our ABC method shows comparable performance as sample size increases in symmetric shape of underlying distribution. However, our method performs much better than other methods when underlying distribution becomes skewed and/or heavy-tailed. The ARE of our method moves towards zero as sample size increase. Some studies applied Bayesian inference to conduct statistical analysis and reported posterior mean and corresponding 95% credible interval. In particular, posterior mean typically does not locate at center of 95% credible interval. In other situation, maximum *a posteriori* probability (MAP) estimate is reported instead of posterior mean. While other existing methods cannot be used for this situation, our ABC method is easily able to obtain standard deviation from these Bayesian summaries. In addition if we only have range or interquartile range and not the corresponding $x_{min}$, $x_{med}$, $x_{Q1}$, $x_{Q3}$, we can use ABC easily to get estimates for means and standard deviation.

## Competing interests

The authors declare that they have no competing interests.

## Authors' contribution

DK and IR conceived and designed the methods. DK conducted the simulations. All authors were involved in the manuscript preparation. All authors read and approved the final manuscript.

# References

1. Wiebe N, Vandermeer B, Platt RW, Klassen TP, Moher D, Barrowman NJ: **A Systematic review identifies a lack of standardization in methods for handling missing variance data**. J. Clin Epidemiol 2006, 95:342-353.
2. Hozo SP, Djulbegovic B, Hozo I: **Estimating the mean and variance from the median, range, and the size of a sample**. *BMC Med Res Methodol* 2005, 5:13.
3. Bland M: **Estimating the mean and variance from the sample size, three quartiles, minimum, and maximum**. *Int J of Stat in Med Res* 2015, 4:57-64.
4. Wan X, Wang W, Liu J, Tong T: **Estimating the sample mean and standard deviation from the sample size, median, range and/or interquartile range**. *BMC Med Res Methodol* 2014, 14:135.
5. Tavaré S, Balding D, Griffith R, Donnelly P: **Inferring coalescence times from DNA sequence data**. *Genetics* 1997, 145(2):505-518.
6. Marin JM, Pudlo P, Robert CP, Ryder RJ: **Approximate Bayesian computational methods**. *Stat Comput* 2012 22:1167-1180.
7. Marjoram P, Molitor J, Plagnol V, Tavaré S: **Markov chain Monte Carlo without likelihoods**. *Proc. Natl Acad. Sci. USA 100,* 2003 15324-15328.
8. Toni T, Ozaki YI, Kirk P, Kuroda S, Stumpf, MPH: **Elucidating the in vivo phosphorylation dynamics of the ERK MAP kinase using quantitative proteomics data and Bayesian model selection**. *Mol. Biosyst.* 2012 8:1921-1929.

**Figure 1**.

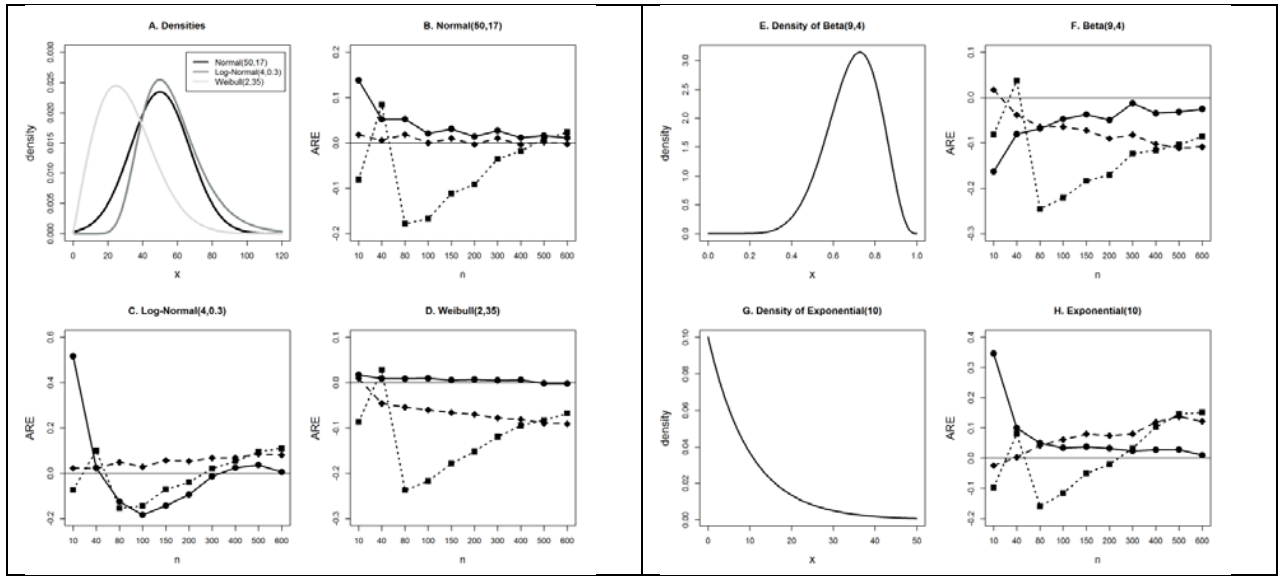

Average relative error (ARE) comparison in estimating sample standard deviation under S1 using simulated data from five parametric distributions. **A, E, G:** Density plots for normal, log-normal, Weibull, beta, and exponential distributions. **B, C, D, F, H:** AREs for 3 methods using simulated data from normal, log-normal, Weibull, beta, and exponential distributions. Hozo et al. (solid square with dotted line), Wan et al. (solid diamond with dashed line), and ABC (solid circle with solid line) methods.

**Figure 2**.

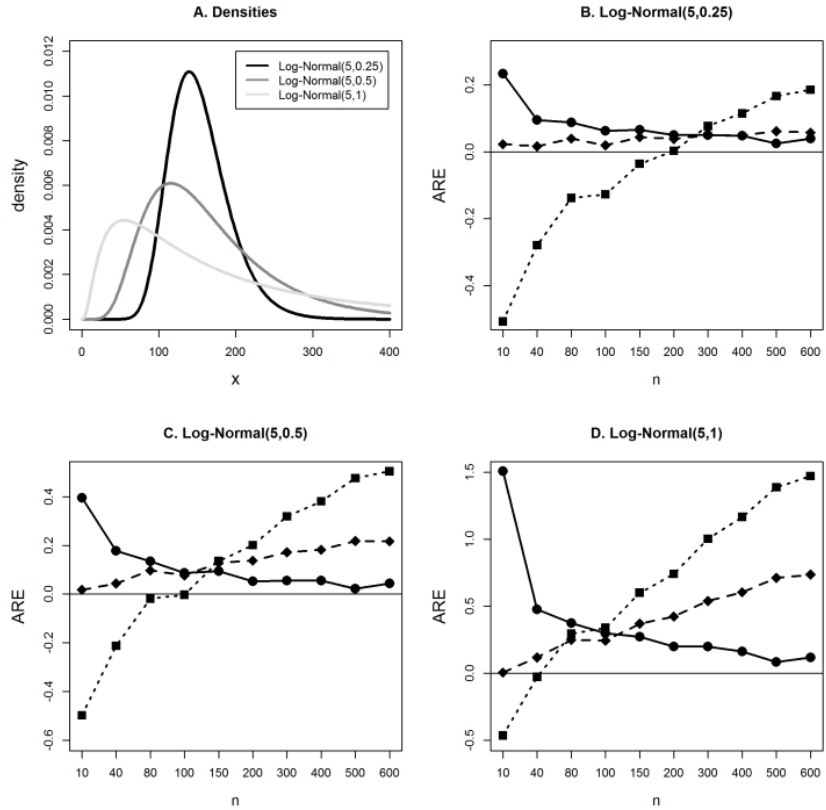

Average relative error (ARE) comparison in estimating sample standard deviation under S2 using simulated data from log-normal distributions. **A:** Density plots for 3 log-normal distributions. **B, C, D:** AREs for 3 methods using simulated data from the same 3 log-normal distributions. Bland (solid square with dotted line), Wan et al. (solid diamond with dashed line), and ABC (solid circle with solid line) methods.

**Figure 3**.

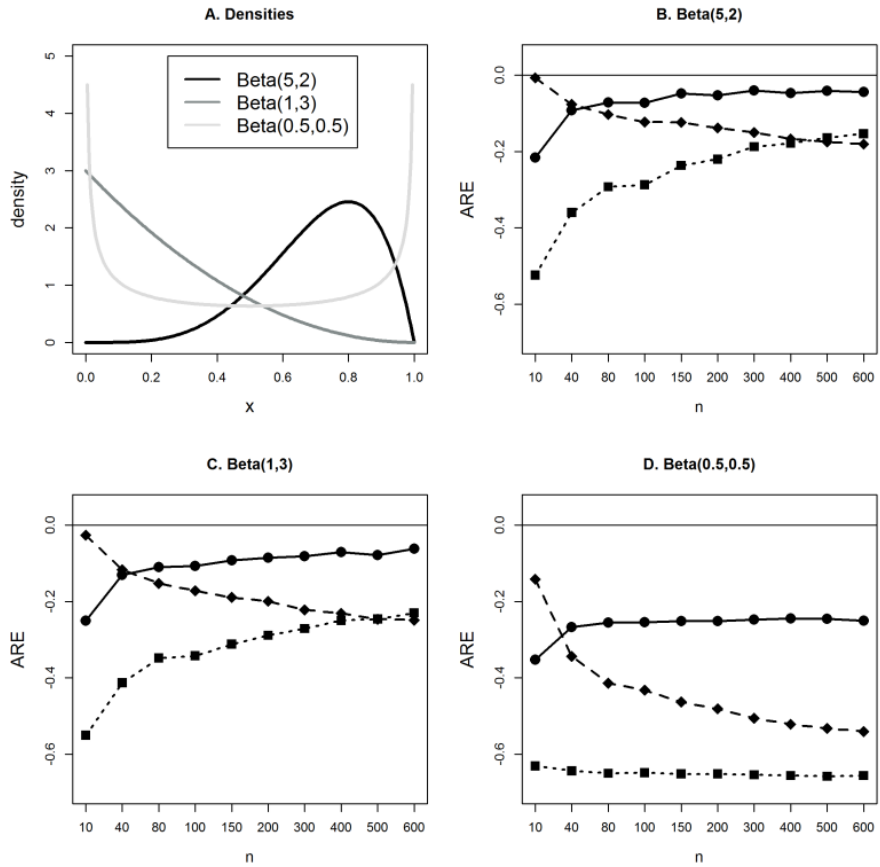

Average relative error (ARE) comparison in estimating sample standard deviation under S2 using simulated data from beta distributions. **A:** Density plots for 3 beta distributions. **B, C, D:** AREs for 3 methods using simulated data from the same 3 beta distributions. Bland (solid square with dotted line), Wan et al. (solid diamond with dashed line), and ABC (solid circle with solid line) methods.

**Figure 4**.

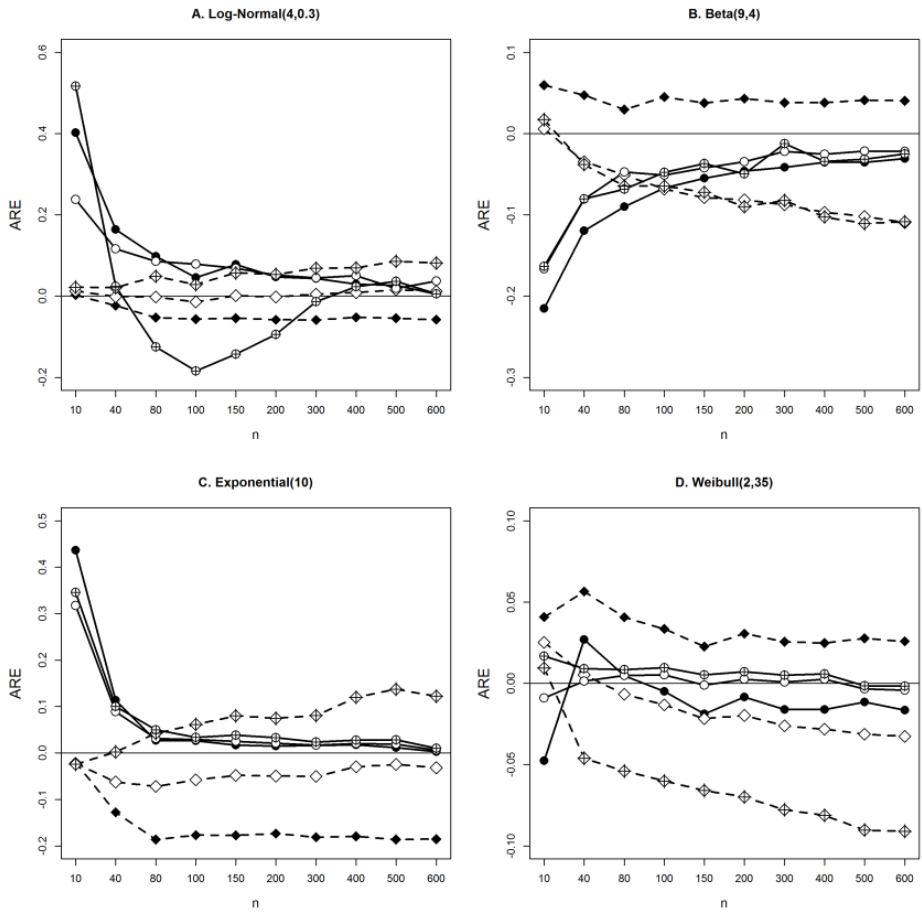

Average relative error (ARE) comparison in estimating sample standard deviation under S1, S2 and S3 using simulated data from four parametric distributions. **A, B, C, D:** AREs for 3 methods using simulated data from log-normal, beta, exponential, and Weibull distributions. Wan et al. (dashed line and crossed diamond for S1, diamond for S2, and solid diamond for S3); and ABC (solid line and crossed circle for S1, circle for S2, and solid circle for S3) methods.

**Figure 5**.

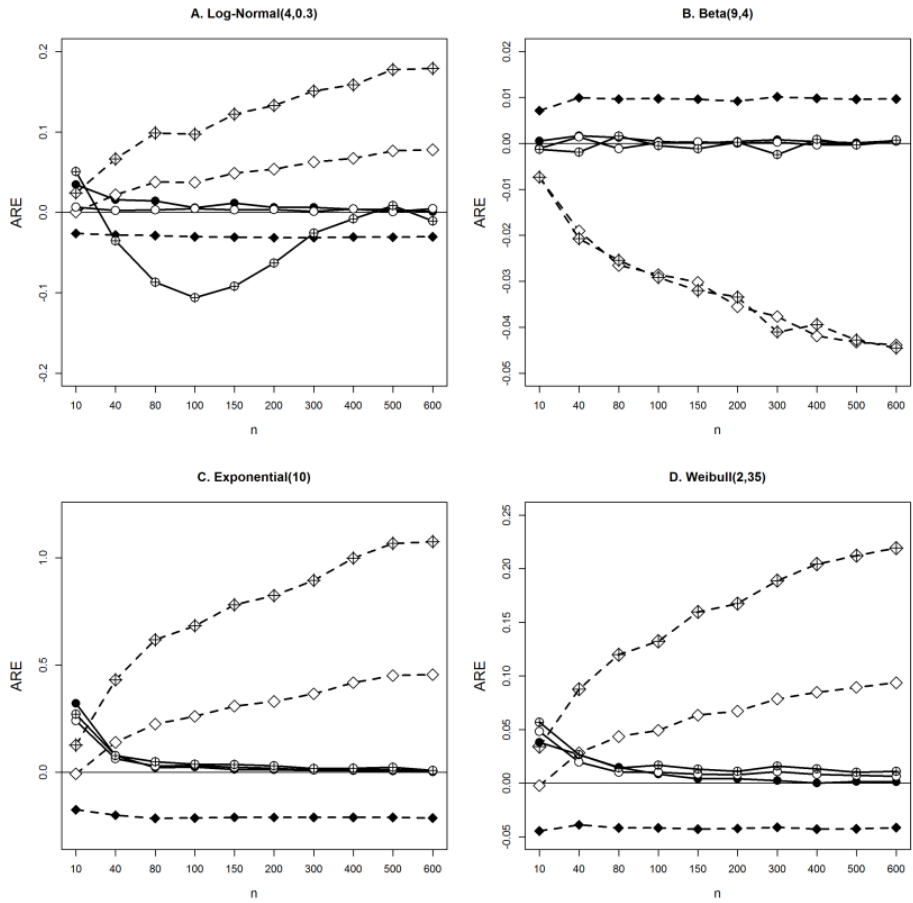

Average relative error (ARE) comparison in estimating sample mean under S1, S2 and S3 using simulated data from four parametric distributions. **A,B, C, D:** AREs for 3 methods using simulated data from log-normal, beta, exponential, and Weibull distributions. Wan et al. (dashed line and crossed diamond for S1, diamond for S2, and solid diamond for S3); and ABC (solid line and crossed circle for S1, circle for S2, and solid circle for S3) methods.

List of Tables

Table 1: Scheme of ABC

| ABC steps | |
|---|---|
| 1 | $\theta^* \sim p(\theta)$; generate $\theta^*$ from prior distribution |
| 2 | $D^* \sim f(\theta^*)$; generate pseudo data |
| 3 | Compute summary statistics, $S(D^*)$, from $D^*$ and compare with given summary statistics, $S(D)$. If $\rho(S(D^*),S(D))< \varepsilon$, then $\theta^*$ is accepted |
| Repeat steps 1-3 many times to obtain enough number of accepted $\theta^*$ for statistical inference | |

Table 2: Priors for ABC in the simulation studies

| Distribution | Parameter 1 | Prior distribution for parameter 1 | Parameter 2 | Prior for parameter 2 |
|---|---|---|---|---|
| Normal (S1) | $\mu$ | Uniform ($X_{min}$, $X_{max}$) | $\sigma$ | Uniform(0,50) |
| Normal (S2) | $\mu$ | Uniform ($X_{Q1}$, $X_{Q3}$) | $\sigma$ | Uniform(0,50) |
| Normal (S3) | $\mu$ | Uniform ($X_{Q1}$, $X_{Q3}$) | $\sigma$ | Uniform(0,50) |
| Log-normal (S1) | $\mu$ | Uniform (log($X_{min}$), log($X_{max}$)) | $\sigma$ | Uniform(0,10) |
| Log-normal (S2) | $\mu$ | Uniform (log($X_{Q1}$), log($X_{Q3}$)) | $\sigma$ | Uniform(0,10) |
| Log-normal (S3) | $\mu$ | Uniform (log($X_{Q1}$), log($X_{Q3}$)) | $\sigma$ | Uniform(0,10) |
| Exponential | $\lambda$ | Uniform(0,40) | - | - |
| Beta | $\alpha$ | Uniform(0,40) | $\beta$ | Uniform(0,40) |
| Weibull | $\lambda$ | Uniform(0,50) | $\kappa$ | Uniform(0,50) |